\documentstyle[prl,aps]{revtex}

\input epsf
\begin{document}

%\draft
\tighten

\def\be{\begin{equation}} 
\def\ee{\end{equation}}

%\input titleabstract.tex
%\input intro.tex
%\input parphys.tex
%\input phasetrans.tex
%\input topdef.tex
%\input cosmoobs.tex
%\input tdcosmo.tex
%\input tdlab.tex
%\input discuss.tex
%\input acknowledgements.tex
%\input bib.tex
%\input bio.tex

%\preprint{\vbox{
%\hbox{-1996}
%}}

\title{Topological Defects in the Cosmos and Lab}
\author{Tanmay Vachaspati}
\address{
Physics Department\\
Case Western Reserve University\\
Cleveland OH 44106-7079, USA.
}

\twocolumn[
\maketitle

\begin{abstract}
\widetext

{\it
Current theories of particle physics lead to the unavoidable
conclusion that there must have been several phase transitions
in the early universe. Further, in the context of these theories,
it is possible that cosmological phase transitions would have produced
topological defects that may be roaming our heavens today.
A finding of these fossils from the early universe would provide
a direct confirmation of the thermal history of the cosmos,
insight into astrophysical phenomena, and, vital information
about particle physics. The elimination of unobserved topological 
defects provides important constraints on particle physics 
and may also suggest novel cosmology. I describe some of the 
research on cosmic topological defects and recent efforts 
to address cosmological issues in condensed matter systems.
}

\

\end{abstract}

%\pacs{}

]

\narrowtext

\section{Introduction}

The theoretical foundations of cosmology were laid by 
Einstein in 1915 with the discovery of General Relativity.
In this framework, it became possible to describe mathematically 
the evolution of the universe and to address questions about its 
beginning and its end. Subsequently, the world of cosmology 
opened up with Hubble's observation that distant galaxies are 
receding, thus leading to the conclusion that the universe is 
expanding. These historic discoveries marked the beginning of 
modern observational cosmology and initiated detailed investigations 
of our universe. Today we can answer questions that earlier we could 
not even imagine asking. 

The observed expansion of the universe means that the younger
universe was smaller and hotter. Using our current knowledge
of physics, this leads to a picture of the universe when it
was only a few minutes old and at a temperature of $10^{10}$ K.
Remarkably, this picture can be (and has been) tested, since the
light elements were ``cooked'' at this time and we can compare calculations
of the cosmological fraction of elements like Hydrogen, Helium, Deuterium, 
and Lithium with their observed abundances. The success of ``Big Bang
Nucleosynthesis'' gives us confidence in our understanding
of the universe from a few minutes after the big bang.

In accelerator experiments, we have studied matter up to energies
corresponding to temperatures of about $10^{15}$ K. The theoretical
description of matter at such temperatures is given by the electroweak 
model due to Glashow, Salam and Weinberg. The triumph of the model was 
in the prediction of the existence of the $W^{\pm}$ and $Z$ bosons which were 
later discovered at CERN. Hence we feel fairly confident
that we understand the behaviour of matter up to $10^{15}$ K. 

The standard model of cosmology that has been so successful in its 
big bang nucleosynthesis predictions, when extrapolated
back to a time of $10^{-10}$ s, predicts that the universe was at 
a temperature of about $10^{15}$ K and so must have been the arena
for electroweak physics. Our confirmation of the electroweak model
provides us with some confidence in our understanding of the
universe at an age of $10^{-10}$ s, though we do not yet have any 
means to directly probe the universe of that time.
At even earlier times, when the universe must
have been at a temperature of about $10^{29}$ K,
particle physicists believe the universe was the stage for the
physics of ``Grand Unified Theories'' (GUTs). Here, we do not yet
have a standard model of particle physics, but there are several
candidates. The exploration of the consequences of particle physics
(and in particular, GUTs) for cosmology, and vice versa, has become 
a subject in its own right.

The electroweak model and GUTs are based on a scheme called ``spontaneous
symmetry breaking'' which, in lay terms, is another name for
phase transitions. If these descriptions of particle physics are
correct, the unavoidable implication is that the early universe 
must have seen phase transitions much like the freezing of water and 
the magnetization of iron. Then, the consequences of phase transitions
that we observe in the laboratory can be expected to apply to 
the universe as well. In particular, relics of the high temperature
phase of condensed matter systems called ``topological defects'' are 
routinely observed in the laboratory and similar relics of the 
early high temperature universe could exist in the present universe.
In other words, these are possible fossils from the early universe.

The hunt for cosmic topological defects depends crucially on their
properties. The last two decades have seen extensive research on 
topological defects and their potential role in cosmology 
\footnote{In recent times, there has been discussion of whether the 
particles that we know ({\it eg.} electrons) are actually topological defects 
\cite{rebbisoliani,tvdual}. This kind of idea has a
long history and the possibility that electrons are objects with 
structure dates back to Abraham \cite{abraham} and Lorentz \cite{lorentz}. 
I will not discuss this very interesting aspect of topological defects 
in the present article.}. 
Very recently, the lack of experimental input has been relieved by enterprising 
condensed matter physicists who have been performing experiments in the 
laboratory to answer questions of great interest to cosmologists. But before 
explaining the possible role of topological defects in the cosmos and the lab, 
I need to describe some basics of modern particle physics.

\section{Inside the Atom}
\label{inatom}

Today, we observe four seemingly different forces in 
Nature. First is the force that holds us on the Earth, 
namely, gravity. Second is the force that keeps the atom
together which is electromagnetism. Then there is the
``weak'' nuclear force which causes radioactivity and the 
``strong'' nuclear force which holds the proton together.

Historically, electricity and magnetism were believed to be
two different forces that were treated in a unified manner only 
once Maxwell wrote his equations. In particular, this means that there 
is only one coupling constant that describes the strength of the 
electric and magnetic forces. Today we understand electromagnetism
as the simplest kind of ``gauge theory''. In fact, the known non-gravitational
forces are ascribed to the exchange of spin 1 particles called gauge 
particles which for the electromagnetic force is none other than the
photon. In mathematical language, the photon is a particle of a
gauge field $A_\mu (t, \vec x )$. Now, it is well-known that there is
a symmetry of electromagnetism related to the gauge transformation:
$$
A_\mu \rightarrow A_\mu ' = A_\mu + 
{1\over e} \partial_\mu \Lambda (t, \vec x )
$$
where, $\Lambda$ is an arbitrary function and $e$ is a coupling
constant. This symmetry is described by rotations in a complex
plane as can be seen if
we couple the photon to a complex scalar field $\Phi$. Then  
an interaction term that preserves the gauge symmetry is
$$
| D_\mu \Phi |^2 \equiv |(\partial _\mu - ie A_\mu )\Phi |^2
$$
provided we also perform the transformation
\be
\Phi \rightarrow \Phi ' = e^{i\Lambda (t, \vec x )} \Phi \ .
\label{gtphi}
\ee
This transformation is a (space-time dependent) rotation in the complex 
$\Phi$ plane, and hence
electromagnetism is invariant under rotations described by one angle
$\Lambda$. Such rotations form a group called $U(1)_Q$ (the group of
unitary $1\times 1$ matrices) where the subscript $Q$ is used to denote 
that the charge associated with the symmetry is ordinary electric charge.

The $U(1)$ symmetry of the model can be ``broken'' or ``hidden'' in 
the vacuum if $\Phi$ takes on a non-vanishing, fixed value in the lowest 
energy state.
This can happen if, for example, there is a potential term for $\Phi$
such as
$$
V(\Phi ) = {\lambda \over 4} ( |\Phi |^2 - \eta^2 )^2 \ .
$$
Then the lowest energy state is obtained with $| \Phi | = \eta$ which
is non-zero, and we say
that $\Phi$ has a ``Vacuum Expectation Value'' (VEV). As the VEV
is not invariant under phase rotations, the $U(1)$ symmetry is said
to be spontaneously broken. Furthermore, by calculating thermal effects
it can be shown that at high temperature, $\Phi =0$ is the lowest energy 
state, while at low temperature $| \Phi | = \eta$ is preferred.
So, if we have a thermal bath of $\Phi$ and $A_\mu$ quanta, at high
temperatures the system will have a $U(1)$ symmetry which will be 
broken upon cooling. This symmetry breaking is depicted as:
$$
U(1) \rightarrow 1 \ .
$$

The reader unfamiliar with group theory might feel lost among the
strange symbols such as $U(1)$ and others to follow. It is best 
to simply think of these as shorthand notations for writing
down all the transformations of the fields in the model that leave
the physics of the system unchanged. So $U(1)$ is just a convenient
way of saying that the transformations that leave Maxwell's equations
unchanged correspond to rotations in the complex plane. Another example,
closer to everyday experience, is the set of continuous transformations 
that leave a sphere unchanged. This is the group of all rotations in 
three dimensions and is denoted by $SO(3)$. 

Using a generalization of the gauge symmetry idea and spontaneous
symmetry breaking, electromagnetism and the weak force have now been 
unified in the Glashow-Salam-Weinberg electroweak model. 
This unification is, however, different from that of 
electricity and magnetism since the unified model still has two coupling 
constants. The unification stems from the fact that the electromagnetic
and weak forces are now described within a common framework.
The electroweak model is based on the gauge symmetry 
\be
SU(2)_L \times U(1)_Y
\label{su2u1}
\ee
which means that the group elements are direct products of 
special (determinant equal to one), unitary, $2 \times 2$ matrices, and, 
phase factors as in (\ref{gtphi}).
The $L$ subscript means that the $SU(2)$ acts on certain
(left-handed) fermions and the $Y$ subscript denotes that the
associated charge is ``hypercharge'' and serves to distinguish
the $U(1)$ symmetry from that of electromagnetism written as
$U(1)_Q$. 
There are four gauge fields in the electroweak model: three ($W^a_\mu$,
$a=1,2,3$) transforming under the $SU(2)_L$ and one ($Y_\mu$) 
under the $U(1)_Y$.

At this stage, it is not evident where electromagnetism is 
contained in the electroweak theory since there is no sign of $U(1)_Q$ in
(\ref{su2u1}). Also, the theory with the symmetry group (\ref{su2u1})
contains four different kinds of massless, spin 1 particles whereas
we only see one (the photon). What happened to three of the four bosons?

Let us now introduce a Higgs (scalar) field, $\Phi$,
which transforms under the group elements in (\ref{su2u1}) and is 
in the doublet representation of $SU(2)_L$ {\it i.e.}
it should be a two complex component vector. 
$\Phi$ is now assumed to
get a ``Vacuum Expectation Value'' (VEV), that is, $\Phi = \Phi_0 \ne 0$.
So now transformations that change the value of $\Phi$ are not allowed.
This means that the symmetry group in (\ref{su2u1}) is no longer valid
and one must find the subgroup that leaves the VEV unchanged. This
subgroup turns out to be a $U(1)$ group and is none other than $U(1)_Q$.
Therefore, after spontaneous symmetry breaking,
there is only one gauge field ($A_\mu$) that is massless just as we
observe, and there are
three gauge fields ($W^{\pm}_\mu$, $Z^0_\mu$) that are massive. 
So the massless photon can mediate long range forces,
while the massive gauge bosons can only mediate short range (weak)
forces. In this way, starting from a very symmetric situation one
derives the vastly disparate electromagnetic and weak forces. 

In this article, I will mainly be interested in the aspect of
spontaneous symmetry breaking 
which in the electroweak model can be depicted as:
$$
SU(2)_L \times U(1)_Y \rightarrow U(1)_Q \ .
$$
In fact, this is not quite correct since the $SU(2)_L$
and the $U(1)_Y$ factors contain two elements that are common.
(This is the center of $SU(2)_L$ which contains the elements
$\pm {\bf 1}$.) So the correct symmetry breaking is:
\be
[ SU(2)_L \times U(1)_Y ]/Z_2 \rightarrow U(1)_Q \ .
\label{ewsymbreak}
\ee
The precise structure of symmetry groups can be very important
in the determination of the cosmological consequences of the
model.

So far we have ignored the strong force.
The theory describing this force is called ``Quantum
Chromo Dynamics'' (QCD) and is based on an unbroken $SU(3)_c$ group 
where the index $c$ stands for the ``colour'' charge. So the standard
model is based on a product of three groups, that is,
$$
[SU(3)_c \times  SU(2)_L \times U(1)_Y ]/(Z_3 \times Z_2) 
$$
With every group there is an associated gauge coupling
constant and so the model has three gauge coupling constants
which are denoted by $g_3$, $g_2$ and $g_1$ for the strong, weak 
and hypercharge factors. 

\begin{figure}[tbp]
\caption{\label{coupcons}
Schematic depiction of the convergence of the three standard model
coupling constants at the grand unification energy scale. The
$g_i$ ($i=1,2,3$) are the various coupling constants in the standard
model, $g_G$ is the GUT coupling constant and $E$ is the energy
at which the coupling constants are measured.
}
\epsfxsize = \hsize \epsfbox{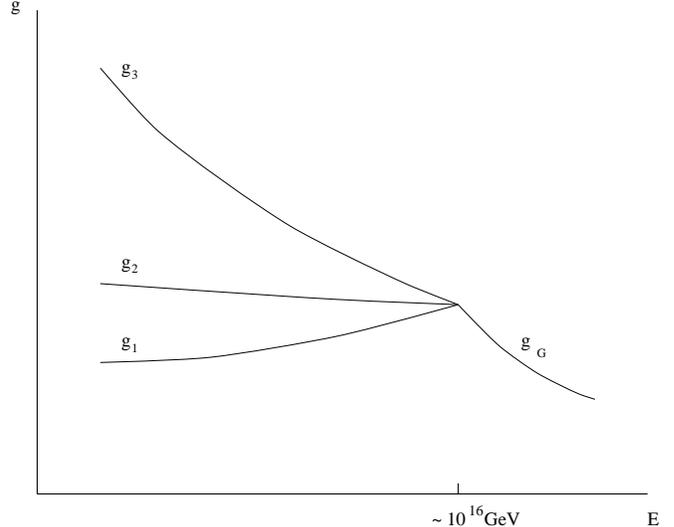}
\end{figure}

In field theory it is known that coupling constants ``run''. This
means that the values of the coupling constants that one measures
depend on the energy at which the measurement is performed. The
rate of the running is determined by the renormalization group
equations which we will not discuss here. But the point is that
the three different coupling constants of the standard model seem
to converge to the same value at an energy scale of about $10^{16}$ GeV 
(see Fig. \ref{coupcons}). This suggests that there is only
one coupling constant at high energies and most likely only one
symmetry group. In other words, the suggestion is that there is
``Grand Unification'' described in terms of a grand unified group.

Let us denote the grand unified group by $G$. Then, as in the
electroweak model, $G$ must break down to the standard model
group which must be a subgroup of $G$:
\begin{equation}
G \rightarrow [SU(3)_c \times SU(2)_L \times U(1)_Y]/(Z_3\times Z_2) \ .
\label{gutsymbreak}
\end{equation}
Two of the simpler examples of $G$ often seen in the literature 
are $SU(5)$ and $SO(10)$.

\section{A Change of Phase}

As we have seen, a central idea in modern theories of 
particle physics is that there is spontaneous symmetry
breaking. However, the idea actually originated in 
condensed matter physics in the context of phase transitions.
To understand the connection between spontaneous symmetry
breaking and phase transitions,
consider a very simple phase transition in which a gas
(or liquid) freezes to form a solid. In the gaseous phase, the 
molecules are flying around in random motion and the
(infinite) container of gas is symmetric under 
translations:
$$
\vec x \rightarrow {\vec x}' = \vec x + \vec \delta
$$
where $\vec \delta$ can be any arbitrary vector. That is,
the symmetry group is that of all translations.
Now, when the gas solidifies, the molecules are arranged
on some lattice and the residual symmetry transformations
are restricted to 
$$
\vec \delta = \vec a
$$
where $\vec a $ is a vector from any one lattice site to another. 
Hence, translational
symmetry has been broken (reduced) by the change of phase.

Going back to particle physics, the very successful electroweak
theory is based on spontaneous symmetry breaking, and hence we
are faced with the prospect of phase transitions in particle
physics. If, somehow, we were to heat up the particle physics 
vacuum, at some high temperature we would be likely to find a new 
phase. For the electroweak phase transition to occur, we expect to
need a temperature of about $10^{15}$ K. (The sun's interior is at 
a mere ten million degrees.) In particle accelerators, we can
achieve the corresponding energies, but only over a very small region
and for a very short duration. So particle accelerators are not
currently useful for studying the electroweak phase transition. 
(They are, however, being used to study the QCD phase transition
at a temperature of $10^{10}$ K.) The GUT phase transition needs an
exorbitant $10^{29}$ K and it would be hard to even dream of a
machine that could attain such energies.
However, the early universe must have seen temperatures corresponding
to the electroweak transition at the age of $10^{-10}$ s and the
GUT phase transition at $10^{-35}$ s, making it the natural environment
for the study of high energy particle physics. At the same time,
particle processes in the early universe must have determined
the state of the current universe and so we would like to 
understand the cosmology of phase transitions. (For a review of
cosmological phase transitions, see the article by M. Gleiser
\cite{mgleiser}.)

An obvious question at this stage is: how can we study something that
happened so long ago? To answer this, I must explain what topological
defects are.

\section{Topology and Frustration}

Let us return to the solidification of a gas. During this
phase transition, the molecules of the gas that are in
random motion have to line up into a regular lattice. If
the gas is cooled quickly, each small volume of molecules
starts lining up but there is not enough time for the 
distant parts of the gas to decide which line to choose.
So molecules in different parts of the gas line up in
a lattice but the orientation of the lattice is chosen
independently. If the orientations are chosen in a certain
way it may become impossible for the entire gas to 
freeze into a regular lattice. This can happen for topological
reasons and the solidification might be frustrated. The end
result is a solid with defects in its lattice. Since these
defects are due to topological conditions, they are known as
``topological defects''. 
(For reviews of topological defects in particle physics and
cosmology, see \cite{avps,mhtk,avpr}.)

\begin{figure}[tbp]
\caption{ \label{wind} 
The winding of the field $\Phi$ around the circle $C$ forces
$\Phi$ to vanish at a point on any surface spanned by $C$. 
By considering different surfaces bounded by $C$, we see that
there is a one-dimensional locus of points
at which $\Phi =0$. Since $\Phi \ne 0$ in the vacuum, there is
energy in the neighbourhood of the curve on which $\Phi =0$. This
energy is locked-in because to remove it, the field would have to
be rearranged over an infinite region of space.
The energy distribution around the curve with $\Phi =0$ is a ``string''.
}

\

\epsfxsize = \hsize \epsfbox{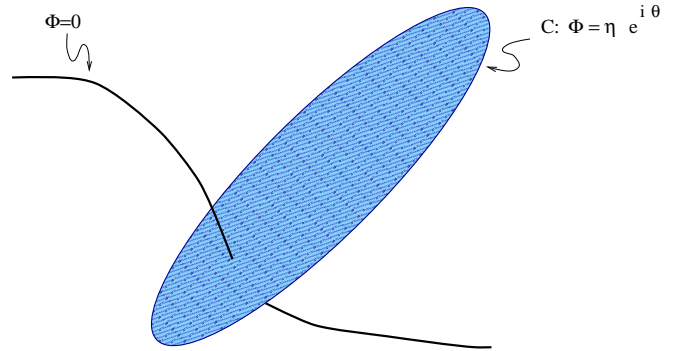}
\end{figure}

To illustrate topological defects in the particle physics
context, consider the $U(1)$ model described in Sec. \ref{inatom}.
Spontaneous symmetry breaking occurs in this model when 
$\Phi (t, \vec x )$ acquires a VEV (that is, becomes non-zero) at
some time. However, as described in a seminal paper by
Tom Kibble \cite{tk},
the acquired value of $\Phi$ at different spatial 
points will, in general, be different. In particular, on a circle $C$
in space, parametrized by an angle $\theta$, we could have:
\be
\Phi \bigr |_C = \eta e^{i \theta } \ , \ \ \  \theta \in [0,2\pi ] .
\label{vevphi}
\ee
There is a topological index associated with this VEV of $\Phi$.
(Basically, it is the number of times $\Phi$ wraps around the circle
in the complex plane as we go around $C$.) Next consider the disk 
bounded by the circle $C$ (see Fig. \ref{wind}). With the value of $\Phi$ 
on $C$ given in (\ref{vevphi}), because of the topology, it is possible 
to show that necessarily $\Phi =0$ somewhere on the disk. But $\Phi =0$ 
is the value of $\Phi$ in the unbroken symmetry phase. Hence the 
completion of the phase transition is frustrated because of the topology 
in the model. Also, the spatial point where $\Phi$ vanishes is not in the 
vacuum (because the vacuum corresponds to $\Phi \ne 0$) and hence, there 
is energy at this point. This energy configuration is called a topological 
defect.

In the case of the $U(1)$ model, we could consider any surface
bounded by the circle $C$, and since $\Phi \ne 0$ everywhere
on $C$, there will always be a point on the 
surface with $\Phi =0$. Therefore there will be a one-dimensional locus 
of points where the phase transition has been frustrated and has $\Phi =0$. 
This one-dimensional topological defect is called a ``string'' and was 
first theoretically described
by Abrikosov in the condensed matter context \cite{abrikosov}, and 
by Nielsen and Olesen in the particle physics context \cite{hnpo}.

Very similarly, phase transitions can get frustrated by topology in
more complicated models. This can result in two-dimensional topological
defects called ``domain walls'', strings with junctions in them, 
point-like defects called ``monopoles'', and, many
hybrids. A distinction is also
made between defects that have associated magnetic fields and those
that have none. The former are called ``local'' or ``gauge'' defects
while the latter are called ``global'' defects. Domain walls are always
global, but strings and monopoles can be global or local. Local monopoles
were discovered independently by `t Hooft \cite{th} and by 
Polyakov \cite{ap}. They are also known as ``magnetic monopoles'' and 
behave just like isolated North or South poles of a bar magnet.
In addition to these topological defects, there is another defect called
a ``texture'' in which the field $\Phi$ is forced to vanish at one point
in space-time.

How can we determine the topological defects in any given model? The secret 
lies in the symmetry breaking pattern which in turn determines the topology
of the vacuum manifold. The point is that, if a certain field configuration
yields the lowest energy state of the system, transformations of this configuration
by the elements of the symmetry group will also give the lowest energy state.
For example, if a spherically symmetric system has a certain lowest energy
value, this value will not change if the system is rotated. More mathematically,
if the group $G$ breaks to a subgroup $H$ (as, for example, in (\ref{ewsymbreak}) 
or (\ref{gutsymbreak})), and the system is in the lowest energy state which we 
denote by $S$, transformations of $S$ by elements of $G$ will leave the energy
unchanged. In addition, transformations of $S$ by elements of $H$ will leave
$S$ itself (and not just the energy) unchanged. So the many {\it distinct} ground
states of the system are given by all transformations of $G$ that are not
related by elements in $H$. This space of distinct ground states is called
the ``vacuum manifold'' and is therefore given by the space of all elements
of $G$ in which elements related by transformations in $H$ have been
identified. The space is denoted by $G/H$ and mathematicians call it a 
``coset space''.

The outcome of the above discussion is that the symmetry breaking leads to
the determination of the vacuum manifold which is some surface in an abstract
mathematical space. Now think of the vacuum manifold 
as a surface like the surface of a ball (two sphere), or, the surface of 
a doughnut (torus). These surfaces have different topological properties. 
For example, one can draw a closed path on a torus that cannot be continuously 
shrunk to a point while all closed paths on a two sphere can be. 
One can also cover the two sphere with another two sphere
(like an orange peel covers the orange)
that cannot then be shrunk to a point. It is these properties that are
crucial for the existence of topological defects.

If the vacuum manifold ({\it i.e.} coset space) has incontractable one 
spheres (paths), the model will have string solutions. (With a little thinking, 
the $U(1)$ example above can help to understand this claim.) If the vacuum manifold 
has incontractable two and/or three spheres, the model contains monopoles 
and/or textures respectively. If the vacuum manifold is disconnected, 
we will get domain wall solutions. The topology of various coset spaces has 
now been determined and is given by what are called ``homotopy groups'' and 
denoted by  $\pi_n (G/H)$. Mathematicians have prepared tables that give
the homotopy groups for different choices of $G$ and $H$.

The basic fact to remember is that the symmetry breaking pattern determines 
the topology of the vacuum manifold and hence the topological defects. 
So given $G$ and $H$ we can determine the topological defects present in the 
system.

%(In the 
%example of the freezing of a gas to a solid, the vacuum manifold is the ``space'' 
%of all possible regular lattices. The topology of this space is important in 
%determining if, in the freezing process, the different parts of the sample can 
%mesh together in a smooth way.) 

An important feature of topological defects is that they cannot be
removed by locally rearranging the fields. In the string case,
for example, the circle $C$ could be chosen to be at infinity and
the removal of the string through the disk would require rearrangement
of the field on an infinite portion of the disk. Any dynamical procedure
to do this would need infinite energy and hence the string is permanently
locked in \footnote{However, if there is a defect and an anti-defect in
the system, they can mutually annihilate.}.

The energy of a defect depends on the temperature at which it
forms. Just to give an idea, monopoles formed at the GUT phase
transition would weigh $\sim 10^{-8}$ gms, strings would have a linear
energy density of about $10^{22}$ gms/cm, and, domain walls would
have a surface energy density of about $10^{52}$ gms/sq-cm.

Not all phase transitions lead to topological defects. A prime example
of such a transition is the electroweak phase transition. (GUT phase
transitions always lead to magnetic monopoles.) Yet it should
be mentioned that there can still be field configurations in the absence
of topology that closely resemble topological defects. Examples of
such configurations include ``semilocal strings'' found by Ana Ach\'ucarro
and me \cite{aatv} and ``electroweak strings'' first found by
Nambu \cite{nambu}. Unlike topological defects, however, these 
configurations are not permanently locked in and can decay.

The possibility of topological defects in particle physics raises
the hope that some of these may have been produced in a cosmological
phase transition and could be observed in the universe today by their
influence on astronomical, astrophysical and cosmological processes.

\section{Cosmological Observations}

\begin{figure}[tbp]
\caption{ \label{lssobs}
The points in the wedges show the distribution of galaxies in a slice of 
the sky as observed by the
Las Campanas Redshift Survey. The survey covers three strips 
of the sky in the Northern hemisphere and another three strips 
in the Southern hemisphere. The larger angular width of the Northern 
hemisphere strips is shown on top of the figure ($10^h$ to $16^h$ of
Right Ascension). The smaller angular widths of the strips is about a
few degrees and the Declination of each of the strips is specified on the 
side of the figure. The radial distance to a galaxy is measured in terms
of a velocity corresponding to the observed redshift of the galactic 
light.
}
\epsfxsize = \hsize \epsfbox{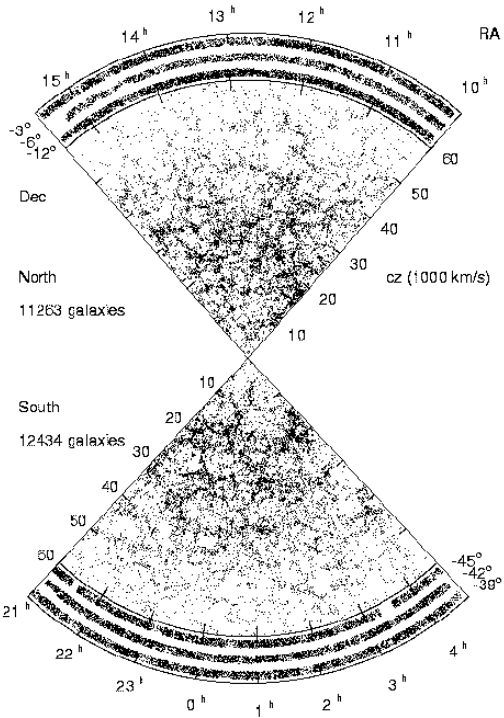}
\end{figure}

Cosmological surveys now cover a large fraction of our observable 
universe.  Astronomers have mapped the luminous structure in slices 
of the sky out to a distance of several hundred megaparsecs 
(see Fig. \ref{lssobs}) \cite{lcrs,cfa}. These maps of 
the universe show that 
galaxies are distributed on walls surrounding empty bubbles (voids). 
This comes as somewhat of a surprise because one's first guess 
would be that galaxies are spread randomly in space.

\begin{figure}[tbp]
\caption{ \label{cmbspec}
The temperature of the CMBR in degrees Kelvin as measured at various 
frequencies in GHz.
(FIRAS was a satellite borne experiment to measure the spectrum of the
CMBR.)
}
\epsfxsize = \hsize \epsfbox{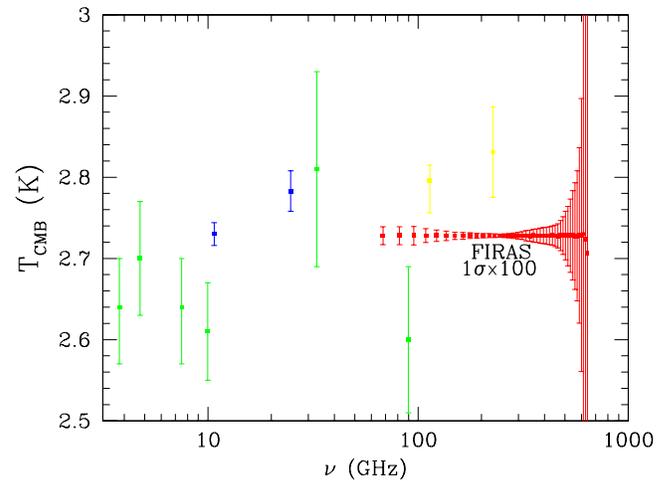}
\end{figure}

Recently, another vital observational tool for the study of the
early universe has become available. This is the structure of the
temperature fluctuations in the
``Cosmic Microwave Background Radiation'' (CMBR). The CMBR is 
light that is directly coming to us from a time when the universe was 
about 100,000 years old and at a temperature of about 3000 K. This
``recombination'' epoch is significant because protons and electrons
combine to form hydrogen atoms at about 3000 K. After recombination, 
the universe contains electrically neutral atoms and, since light 
does not scatter off neutral atoms, it can travel freely to us. Before 
recombination, however, 
the matter in the universe is electrically charged and light scatters
strongly. During this period, light propagates as if it were in a 
fog and so light from the pre-recombination universe cannot reach us. 
The CMBR is the earliest light we could possibly see and it is very
significant that we have actually seen this light 
(see Fig. \ref{cmbspec}) \cite{firas}.

\begin{figure}[tbp]
\caption{ \label{cmbcls}
The observed distribution of $<Q>$ - a quantity related to the anisotropy
of the CMBR in the $l^{th}$ multipole moment (see text) - together with 
error bars. The
curve is the prediction of an inflationary model. (The $<Q>$ in this
plot is normalized with an extra factor of $(5/12)^{1/2}$ as compared 
to the definition in the text for historical reasons.)
}
\epsfxsize = \hsize \epsfbox{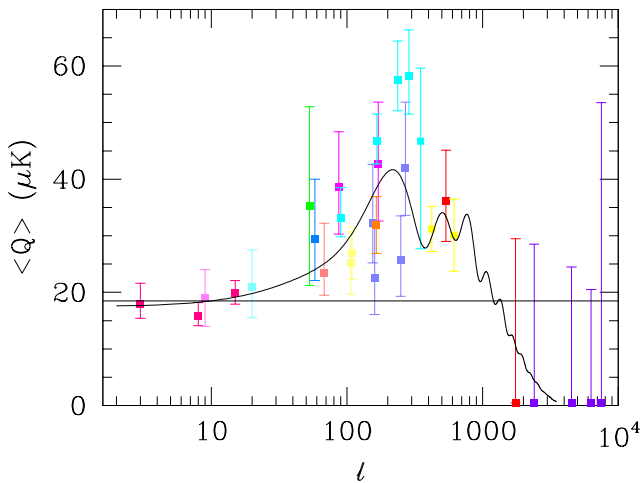}
\end{figure}

The CMBR is extremely uniform in all directions. The uniformity
is only spoilt by 
tiny fluctuations of about 1 part in $10^5$. In other words, the
temperature of the CMBR is $T=2.7$ K, no matter in which direction
you choose to look but there are fluctuations $\delta T$ in this
temperature:
$$
\left ( {{\delta T}\over T} \right )_{rms}  \simeq 10^{-5} \ .
$$
Further, due to the growth in the number of observational experiments,
it is now becoming possible to say something about the map of
$\delta T$ over the sky. The observations determine the
temperature fluctuation on the sky at different angular scales and so
one has quantities related to the spherical harmonics of $\delta T$.
The usual procedure in a calculation of the anisotropy is to decompose 
the temperature fluctuations on the sky 
(coordinates $\theta$ and $\phi$) in spherical harmonics:
$$
{{\delta T} \over T} (\theta , \phi )
= \sum_{l=0}^\infty \sum_{m=-l}^{m=+l} a_{lm} Y_{lm} (\theta , \phi )
$$
and then calculate 
$$
c_l = < | a_{lm}^2 | > ~ ,
$$
where the angular brackets denote an ensemble average. Then it is
conventional to find
$$
<Q> \equiv \left [ {l(l+1)c_l}\over {2\pi} \right ]^{1/2} T_{cmbr}
$$
as a function of $l$,
where $T_{cmbr}$ is the CMBR temperature measured in $\mu$K.
The calculated $<Q>$'s are then compared with observations
(see Fig. \ref{cmbcls}) \cite{cobeandmore}.

The fluctuations of the temperature of the CMBR can be produced in
two ways. The first is if the matter that the photons last scattered
off (the ``last scattering surface'') was not quite uniform, and the
second is if there are objects between the last scattering surface
and the present that disturb the photons. In the first case, the CMBR
provides a very definite determination of the state of the 100,000 year
old universe and in the second case, it provides a probe of the universe
between now and the last scattering surface. By considering the details
of the fluctuations of the CMBR, we hope to be able to derive both the
state of the early universe and the intervening influences.

\section{Fossils from the Early Universe}

Based on our knowledge of particle physics, the
gradual cooling of the universe must have been
punctuated by sharp phase transitions. This is
similar to the violent climatic changes on earth
that would have affected the otherwise gradual 
evolution of life. Further, just as we seek fossils
of the early forms of life, we can seek fossils from 
the early universe in the form of topological defects.
In fact, topological defects are our main hope of
directly studying the very early universe.

The current belief that the electromagnetic, weak and 
strong forces unified at about $10^{16}$ GeV implies
a cosmological GUT phase transition at a temperature
of $10^{29}$ K at the young age of $10^{-35}$ s. Then
we are led to consider the formation of topological
defects corresponding to this scale. These defects
could be magnetic monopoles, strings or domain walls.

Magnetic monopoles formed at the GUT scale would 
dilute with the expansion of the universe while keeping
their number fixed. This means that the energy density in
monopoles goes down as $a^{-3}$ where $a$ is the scale factor 
of the universe. However, the dominant energy in the 
early universe is radiation. The energy density of radiation
not only gets diluted by the expansion but the energy of
each radiation quanta also gets red-shifted. Therefore
the energy density in radiation falls off as $a^{-4}$. This
means that the energy in monopoles becomes more important
as the universe expands. Following this argument by a more
careful and detailed analysis, Zel'dovich and Khlopov \cite{yzmk},
and John Preskill \cite{jp} found 
that GUT monopoles would start dominating the universe very early 
and would overclose the universe ({\it i.e.} the energy density 
in monopoles would exceed the critical density and the universe 
would recollapse in a very short time). This is clearly not the case.

Around 1980, the monopole overabundance problem led to a tension 
between a central belief in particle physics - that of grand 
unification - and cosmology. For consistency, either grand unification 
had to succumb, or, cosmology needed revision. The breakthrough was
achieved when Alan Guth realized \cite{ag} that an exponential inflationary 
period in cosmology, during which the energy density in monopoles is 
diluted to acceptable levels, would alleviate the tension. 
The following years have seen a number of other solutions 
to the monopole problem but inflationary cosmology has survived because 
it also offers solutions to a number of other cosmological puzzles.
(See Andrei Linde's book on inflationary cosmology \cite{linde} for an
account of the field.) 

Domain walls formed at the GUT scale, like magnetic monopoles, would 
be a cosmological disaster. If we had one domain wall of mass per unit
area equal to $\sigma$ in our visible universe (size $t$), the total
energy contained in it would be of order $\sigma t^2$. And the energy 
in all the other matter would be of order $\rho t^3 \sim t/G$ where
$\rho \sim 1/ Gt^2$ is the energy density in matter and $G$ is
Newton's gravitational constant. So the ratio of domain wall energy
to other forms of energy is $G\sigma t$. By inserting the value
of $\sigma$ ($10^{52}$ gms/sq-cm) and $G$, we find that domain walls
start dominating the universe very early and would lead to a universe
unlike ours. This rules out the formation of GUT domain walls. 
(Though, if the GUT model leads to an inflationary universe,
GUT domain walls would be acceptable.)

Cosmic strings formed at the GUT scale are more benign than magnetic
monopoles and domain walls. To see this, however, is considerably
more difficult. The problem has been studied over the years
using intensive computer simulations by
three groups: Andy Albrecht and Neil Turok, 
Dave Bennett and Francois Bouchet, and 
Bruce Allen and Paul Shellard (see \cite{proceed} for reviews). 
Analytic tools to study the problem have been devised by
Tom Kibble, Dave Bennett, Ed Copeland and others (references may be
found in \cite{mhtk,avps}). Most recently, Mark Hindmarsh, inspired by 
ideas from condensed matter physics, has devised a technique to
study the evolution of domain walls in an expanding universe \cite{mh}. 
This seems like a 
promising approach to study the evolution of cosmic strings too, though
the technique has not yet been applied to this problem.

The results on cosmic string evolution, first
established by Bennett and Bouchet, were that the network
evolves in such a way that the long strings shed very small loops and
the energy density in strings remains a fixed
small fraction of the energy density in the universe. This leads
to the possibility that cosmic strings may have been produced at the
GUT scale and could be ``out there'' for us to discover. In addition,
cosmic strings would have influenced the light and matter around them
and so it may be possible to detect this influence by careful observations
of the present universe. In particular, cosmic strings may have left
their imprint on the CMBR and the large-scale structure. 
The flip side of the coin is that if the effect 
of GUT scale cosmic strings on the CMBR 
and on large-scale structure formation disagrees with observation,
we would be able to say that they do not exist and thus gain some
important information about particle physics at very high energies. 

\begin{figure}[tbp]
\caption{ \label{lens1}
The filled circles show the location of several hypothetical unlensed
sources in the presence of a foreground string segment that was generated
by computer simulation. The sharp kinks in the string
are partially due to the fact that what is shown is the projection of the
string onto a plane and, on small scales, due to the simulation grid used
to generate the string.  The inner box (dotted line) is 25''$\times$25''.
}
\epsfxsize = \hsize \epsfbox{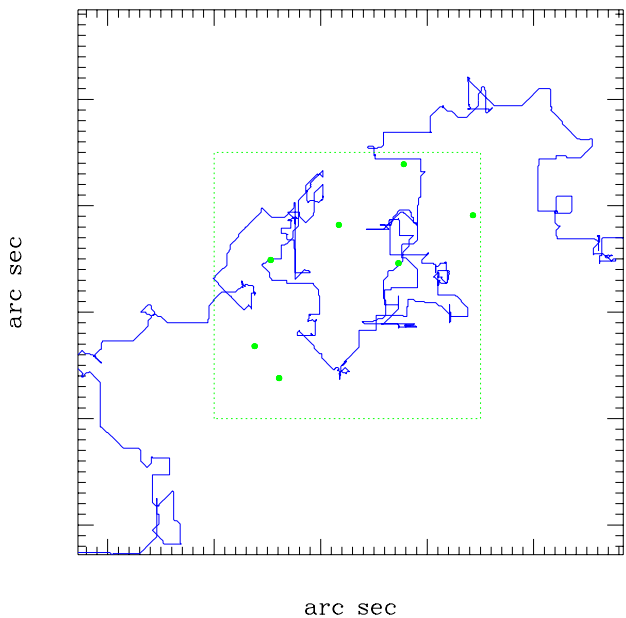}
\end{figure}

\begin{figure}[tbp]
\caption{ \label{lens2}
The appearance of the field of sources in the 25''$\times$25'' size
box shown in the previous figure due to gravitational lensing by the
string. The stringy appearance of the lensed sources seems evident.
The challenge in real surveys would be to pick out the stringy nature
of the signal in a field of other astronomical objects.
}
\epsfxsize = \hsize \epsfbox{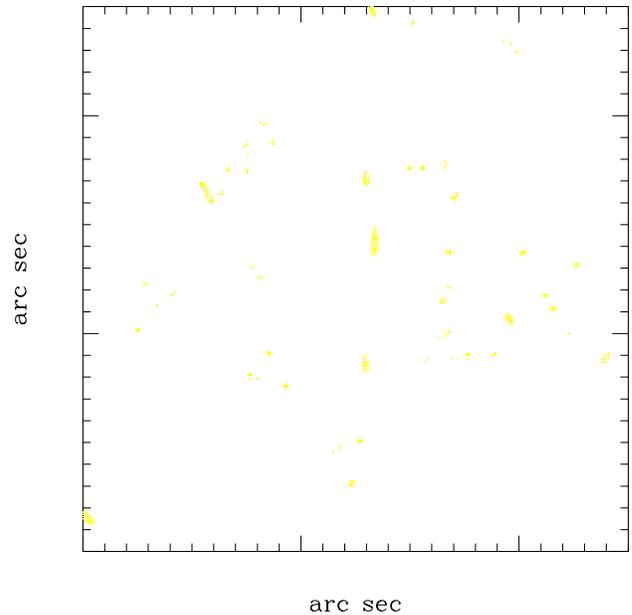}
\end{figure}

There are two known ways to hunt for cosmic strings. The first is by
realizing that a cosmic string that is illuminated on the backside by 
a light source would act as a lens for the source since the string
curves the intervening spacetime. So a string would cause multiple 
images of a background quasar or galaxy. The observation of such an
event would not only tell us that there are cosmic strings in the universe
but it would also tell us where the string is currently located. 
With Andrew de Laix and Lawrence Krauss, I have recently been investigating 
this scheme for a cosmic string hunt \cite{adlktv}. 
In Fig. \ref{lens1} the location of a 
string with several background sources is shown. The string causes the light 
from the sources to bend and the field appears as shown in Fig. \ref{lens2}. 
In this hunt for strings, there is an uncertainty in the details of the 
lensing pattern since the shape of cosmic strings is not precisely
known. However, the 
limiting factor is the small probability for looking in the right direction 
for observing a string lensing event. Ongoing and planned surveys, however, 
will be covering roughly a quarter of the sky and should find GUT cosmic 
strings if they are there.

A second way to search for strings is to seek their imprint on the
CMBR. Just as a string distorts the
images of background sources, Kaiser and Stebbins \cite{nkas}
showed that moving strings would change the energy of photons
that pass by. Since the CMBR
is background illumination for cosmic strings, it should have
temperature fluctuations induced by strings. Ongoing experiments are
determining the CMBR fluctuations very
carefully and theorists have been calculating the influence of strings
on the background. The theoretical predictions depend on various other
factors (such as the energy density in the universe). At the moment,
the simplest cosmological model with strings does not appear to be
consistent with the observations \cite{ballenetal,aalbrechtetal}.
In another 5 years, with more and better data and with further
characterization of the string network, we should be able to say with
greater certainty if the observed anisotropy in the CMBR can be due to
GUT cosmic strings.

In addition, as first pointed out by Zeldovich \cite{yz} 
and by Alex Vilenkin \cite{avlss}, it is possible to consider the influence 
of cosmic strings on the formation of structure (galaxies {\it etc.}) in the 
universe. Over the years, 
our understanding of the influence of cosmic strings 
on the matter in the universe has evolved. At first it was believed that 
cosmic string {\em loops} would be centers around which galaxies would
form. Later the potential importance of long (infinite) strings for structure 
formation was realized by Silk and Vilenkin \cite{jsav}, 
Stebbins {\it et. al.} \cite{asetal} and by me \cite{tvlss}. 
This realization
gained force once it was found that the string loops were too small to be
of much importance in the formation of large-scale structure \cite{tvav}.
The implications of cosmic strings for structure formation continue to
be worked out with great vigour by researchers like Albrecht, Allen, 
Brandenberger, Shellard, Stebbins, and others. 

It should be added that there are great cosmological, astrophysical and 
theoretical uncertainties in the research on formation of large-scale structure
by strings ({\it eg.} see the paper by Rees \cite{rees}). 
However, if the calculated distribution of large-scale structure due to
strings roughly agrees with the observed distribution 
(Fig. \ref{lssobs}), this would provide hope for the existence of strings. 
A disagreement would provide evidence against string seeded structure 
formation and hence a constraint on GUT models in particle physics. 

As first pointed out by Hogan and Rees \cite{hoganrees}, there
is yet another constraining observation that cosmic strings must
satisfy - this is the observed limit on a cosmological background
of gravitational waves. Since cosmic strings generate gravitational
radiation, their energy density has to be low enough such that their
gravitational radiation remain within limits imposed by the timing
of the millisecond pulsar \cite{jtaylor}. 
(A gravitational wave background would introduce noise in the 
millisecond pulsar timing 
beyond that what is observed and accounted for.) An estimate of the 
gravitational radiation from strings depends sensitively on the 
structure of the string network. Based on the current understanding of
the network, the gravitational wave constraints are evaded
by GUT strings, though by a small margin.

In an effort spearheaded by Turok \cite{turok}, 
Spergel \cite{spergel} and Durrer \cite{durrer}, cosmologists 
have also examined the influence of texture 
and other global defects on the CMBR and large-scale structure. Once again,
analysis of the simplest theoretical models indicates that GUT scale
global defects by themselves cannot simultaneously explain
large-scale structure formation and the anisotropy of the CMBR.

The interest in the GUT phase transition comes from the underlying
unification philosophy, the apparent convergence of the known 
coupling constants (Fig. \ref{coupcons}), 
and, the cosmological relevance of the GUT
energy scale. (The GUT energy scale seems suitable for laying out
the seeds of density inhomogeneities that will later grow to become 
galaxies.) However, our knowledge of particle physics is not yet 
complete enough that we can say that the electroweak and GUT phase
transitions were the only unifying cosmological phase transitions. 
Indeed, there are several particle physics models in which phase
transitions would have occurred between the electroweak and GUT
epochs. Defects produced at these epochs may not have been responsible
for galaxy formation but it would be invaluable to know if they
exist in the universe. Since these defects would be lighter, it
is unlikely that they will be seen due to their gravitational interactions. 
Instead, to hunt them, one has to rely on their particle physics
interactions which can lead to electromagnetic radiation and cosmic
rays, an effort actively pursued by Bhattacharjee and others 
\cite{pbgsds,tkay,avcosmicrays}.

\section{Down on Earth}

The fact that cosmological phase transitions and condensed matter 
phase transitions are described by the same physical principles, 
allows us to consider performing ``cosmological experiments''
in the lab. These are experiments in condensed matter systems that 
are motivated by cosmology. 
This idea was first suggested by Zurek \cite{zurek}. For example,
condensed matter physicists have studied topological defects for 
a long time and have been interested in their microphysical properties 
and also in how the system of defects relaxes
with time thus leading to the completion of the phase
transition. However, until now, they were not
interested in the number of defects, or in the size
distribution of vortices (strings) that are formed during
a phase transition. Both these quantities are of crucial interest
to cosmologists since the number and distribution of
defects determines their astronomical and astrophysical
relevance. Hence an experiment that studies the 
distribution of vortices produced during a phase transition
would be called a ``cosmological experiment'' \footnote{The exchange of 
ideas between cosmologists and condensed matter physicists was greatly 
facilitated by a six month long program and a NATO workshop on topological 
defects held at the Isaac Newton Institute during 1994. The workshop
lectures can be found in the proceedings edited by Davis and
Brandenberger \cite{robertanne}.}.

Over the last several years, a number of condensed matter
experiments of a cosmological flavour have been performed. First
was the experiment in nematic liquid crystals by 
Chuang {\it et. al.} where the authors studied the relaxation 
of a network of strings \cite{chuang}. This was later followed
by efforts to study the formation of defects in liquid crystals
by Srivastava and collaborators \cite{ams}. 
The attention then turned to phase 
transitions closer to expected particle physics phase transitions 
and experiments studying the formation of vortices in $^4$He
were carried out by Peter McClintock and his group in Lancaster
\cite{he4}. Most recently, there have been a number of ingenious 
experiments in $^3$He conducted in Grenoble, Helsinki and
Lancaster \cite{grenoble,helsinki}
that have also studied the formation of strings. 
(These are described in the article by A. Gill \cite{agill}.) 

A leading personality in the effort to connect particle physics and
cosmology with $^3$He is Grisha Volovik. The point he has tried to
convey to the physics community is that there are strong similarities 
between the basic structure of particle physics and $^3$He \cite{gvtv}. 
So one can imagine simulating particle physics processes of cosmological 
interest in $^3$He provided one is careful to ask the right questions. 
For example, as has been done with astounding success, it is possible to 
simulate the cosmological formation of strings in $^3$He since the formation
of topological defects is not peculiar to details of the cosmological 
environment. At the same time, it seems that it may not be
possible to simulate the cosmological evolution of strings in condensed
matter systems since that depends on the 
Hubble expansion and the absence of strong
dissipative processes, both of which are cosmological conditions and
hard to find in the lab setting.
Here I will describe another process that has been studied in $^3$He
\cite{baryo3he}
and which is of great interest to cosmologists - this is the generation
of matter, also called ``baryogenesis''.

In the absence of an external magnetic field, 
$^3$He is known to have two superfluid phases which are
called the A and B phases. At high temperature, 
$^3$He is invariant under rotations of the Cooper pair
spin (S), orbital angular momentum (L), and, also, phase 
rotations of the wave-function that lead to the
conservation of particle number (N). So the (continuous)
symmetry group of $^3$He is:
$$
G = SO(3)_S \times SO(3)_L \times U(1)_N \ .
$$
The spontaneous symmetry
breaking pattern for the transition into the A phase is:
\be
G \rightarrow SO(2)_{S_3} \times U(1)_Q \ ,
\ee
where 
$$
Q = L_3 - {N \over 2}
$$
Note that in this symmetry breaking, the $SO(3)_S$ group
breaks to $SO(2)_{S_3}$ while 
the remaining symmetry breaking pattern appears to be exactly 
that of the electroweak model. There are, however, subtle
differences in certain discrete symmetries in the two
models that are important in determining the topology 
of the vacuum manifold and hence, the topological defects.
(Nonetheless, a direct analog of the non-topological electroweak 
string is present in $^3$He \cite{gvtv}.) Another difference is 
that $^3$He does not contain fundamental gauge fields other than the 
electromagnetic fields.  
In the electroweak model, however, such gauge fields exist and 
are important.  It is useful to be aware of these subtle differences 
because it allows us to meaningfully compare $^3$He experiments with
particle theory expectations.

The common elements in $^3$He and (hypothetical) particle theory 
is the presence of non-trivial topology. Therefore processes
such as the formation of topological defects can be studied in
$^3$He and the results translated to the particle physics world.
In addition, $^3$He contains quasiparticles that correspond to the 
fundamental particles (leptons and quarks) in particle theory. 
These quasiparticles interact with the order parameter of $^3$He 
just as the fundamental fermions interact with the electroweak gauge 
fields. So $^3$He does contain``effective'' gauge fields besides the 
ordinary electromagnetic fields.
This similarity is very valuable since the behaviour of fermions
in fixed background gauge fields can be simulated 
by the interaction of quasiparticles in the background of some
order parameter configuration in $^3$He. 
Indeed this is precisely what is needed
to simulate the violation of baryon number in $^3$He.

\begin{figure}[tbp]
\caption{\label{zeromode}
{\bf a)}
The energy versus the momentum of fermions along the $^3$He vortex
(assumed to lie along the $z-$axis). The crucial feature in this
spectrum is the presence of a zero mode (the $n=0$ line) which
crosses from the negative energy to the positive energy region. The
application of an electric field lifts the level of
the Dirac sea along the $n=0$ mode and particles move from the
vacuum ($E<0$) to the physical world ($E>0$).
This is the anomalous production of fermions from the vacuum.
{\bf b)} The spectrum of $u$ and $d$ quarks on strings in the
electroweak model. As the level of the Dirac sea rises (or falls),
$u$ and $d$ quarks are produced (or destroyed) in such a way that
the total electric charge, $q$, remains conserved but the total
baryon number $B$ is violated. ($C$ is the charge under an
operation called ``particle conjugation''.)
}
\epsfxsize = \hsize \epsfbox{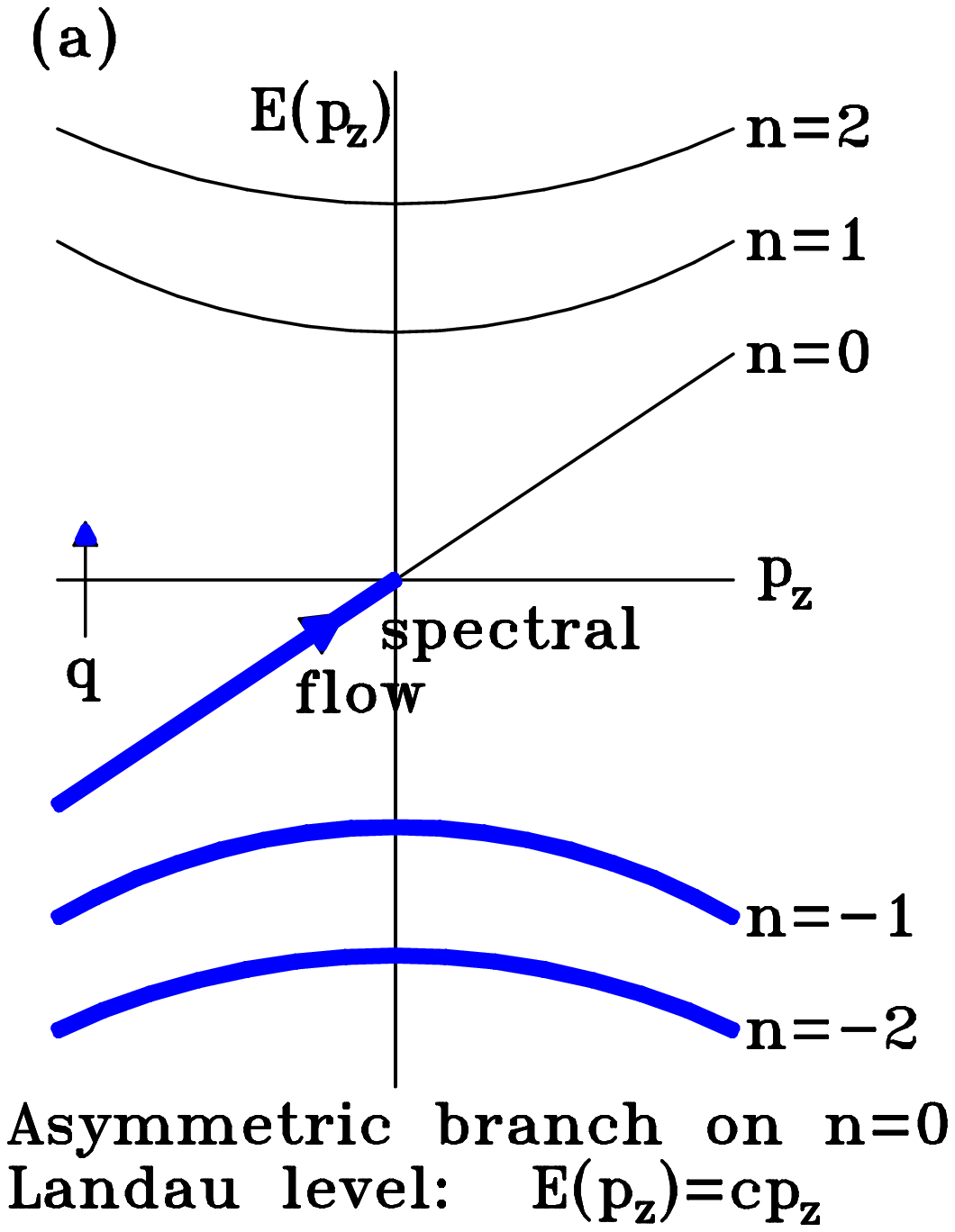}
\epsfxsize = \hsize \epsfbox{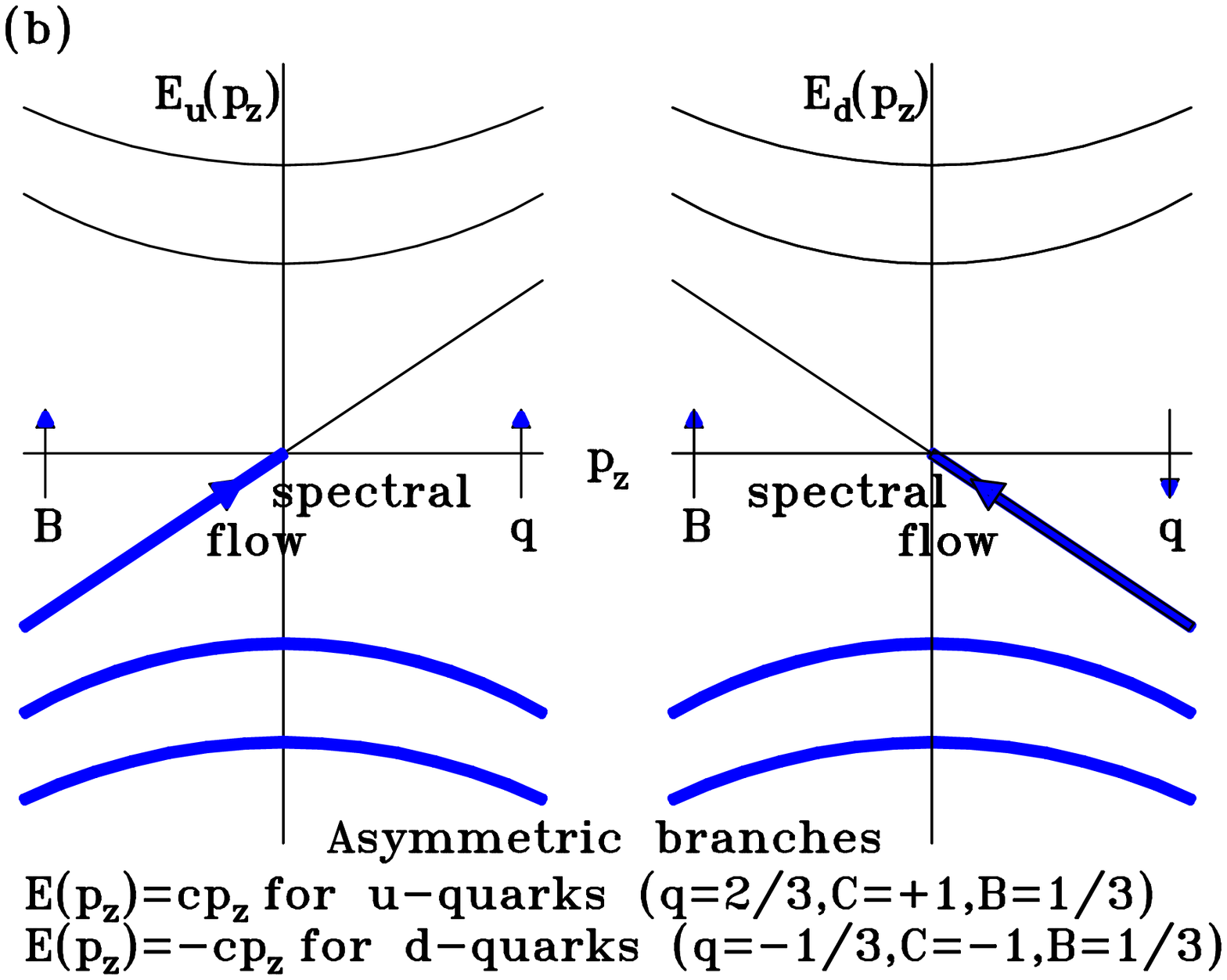}
\end{figure}

In vacuum the energy of a free fermion is given by:
$$
E = \pm \sqrt{{\bf p} ^{~ 2} + m^2 } \ ,
$$
where $\bf p$ is the momentum and $m$ is the mass of the
fermion ($c$ has been set to 1). Therefore to create a fermion
and antifermion pair from the vacuum requires at least an
energy equal to $2m$.
In the presence of certain scalar and gauge field configurations,
however, the dispersion relation for fermions can display a
``zero mode'' (see Fig. \ref{zeromode}). 
This can be seen by solving the Dirac equation in the non-trivial
scalar and gauge field background. If the Dirac equation has a
solution with zero energy eigenvalue then this solution is the
zero mode. (Alternatively, the existence of the zero mode follows 
from certain mathematical ``index theorems'' which I will not
describe here.) For example, there can be a zero mode in the background 
of a string lying along the $z-$axis. Effectively this says that
there are fermions trapped on the string that behave as if they are 
massless as long as they stay on the string. But these trapped fermions
can travel along the string and their dispersion relation in the
one dimensional space of the string is:
$$
E = \pm ~ p_z   \ .
$$

The existence of zero modes leads to the anomalous creation
of fermions (Fig. \ref{zeromode}). An intuitive picture
is that, if an electric field is applied along the string,
the Dirac sea can rise as a whole and particles from the Dirac
sea can get pushed into positive vacuum states along the zero mode. 
This can happen in string-like configurations in the electroweak model 
of particle physics and also along vortices in $^3$He. The anomalous 
creation of fermions in the electroweak model leads to the creation of
matter (baryons) over antimatter (antibaryons) or vice versa,
while the anomalous creation of quasiparticles in $^3$He 
leads to the violation of total quasiparticle momentum and is
observed as an excess force on moving vortices. 
In a cosmological scenario, such processes
together with other suitable conditions such as thermal non-equilibrium
and CP violation can lead to the creation of matter in the universe
(baryogenesis). 

When we apply an electric field ${\bf E}$ along a string that
carries magnetic field ${\bf B}$, the
rate of production of fermions of charge $q$ is:
\be
\dot n = {{q^2} \over {4\pi^2}} {\bf E}\cdot {\bf B} \ ,
\label{ndot}
\ee
where $n$ is the number density of fermions.
(The electric field itself can be induced via Faraday's law if 
the string moves across an ambient magnetic field.) 

\begin{figure}[tbp]
\caption{\label{link}
A knotted configuration of electroweak strings that has associated
baryon number. Here ``baryon number'' is defined in terms of
particles trapped on the string and this is somewhat different from
the usual meaning which is defined in terms of particles in the vacuum.
}
\epsfxsize = \hsize \epsfbox{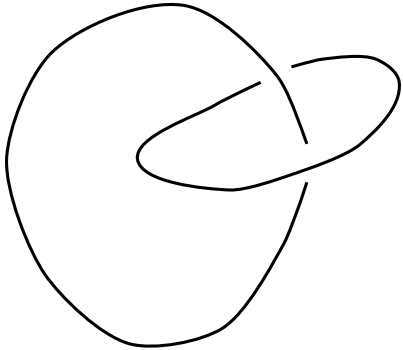}
\end{figure}

The anomaly equation (\ref{ndot}) is applicable to both the electroweak
model and $^3$He. The possibility of anomalous generation of baryon number 
along strings was discussed by Witten \cite{ew}. 
In the electroweak case, a non-Abelian generalization
of (\ref{ndot}) leads to the possibility of anomalous baryon charge
on electroweak string knots (see Fig. \ref{link}) as I showed in 
collaboration with George Field \cite{tvgf}, and, Jaume Garriga \cite{jgtv}. 
In $^3$He, the 
anomaly equation leads to quasiparticle production. The measurable quantity, 
however, is the momentum, $\bf P$, carried off by the anomalously created 
quasiparticles:  
$$
\partial_t {\bf P} = {{1} \over {2\pi^2}} 
          \int d^3 x (p_F {\bf {\hat l}}) {\bf E} \cdot {\bf B} \ ,
$$
where, $p_F$ is the Fermi momentum and ${\bf {\hat l}}$ is the orientation
of the Cooper pair angular momentum.

In the Cooper pair plus quasiparticle system, momentum is obviously
exactly conserved. In the absence of the anomaly, the momentum in
the Cooper pairs and quasiparticles is separately conserved. Due to
the anomaly, however, momentum is transferred from the $^3$He vacuum
(Cooper pairs) to the quasiparticles and vice versa. This transfer
of momentum leads to an extra force on moving vortices:
\begin{equation}
{\bf F} = \partial_t {\bf P} = \pi \hbar N C_0 
           {\bf {\hat z}} \times ({\bf v}_n - {\bf v}_L ) \ ,
\label{extraforce}
\end{equation}
where $N$ is the winding of the vortex, the coefficient $C_0$ is
a temperature dependent coefficient,
the vortex lies in the ${\bf {\hat z}}$ direction, and
${\bf v}_L - {\bf v}_n$ is the vortex line velocity with respect to
the normal fluid. 

The Manchester group, led by Henry Hall and John Hook, 
used a clever experimental setup 
in which an array of vortices was created by rotating a sample of $^3$He. 
A diaphragm placed within the sample had two orthogonal modes of oscillation
which could be driven electrically and also detected. Oscillations in one of 
the modes was used to create the relative velocity ${\bf v}_L - {\bf v}_n$.
The extra force on the vortices given by eq. (\ref{extraforce}) produces
forces perpendicular to the driven mode of oscillation and hence couples 
to the other oscillation mode of the diaphragm. The oscillations in this
orthogonal mode can then be measured. This leads to the measurement of 
quantities related to the coefficient $C_0$ at different temperatures. 
The results confirm the anomalous production of quasiparticles 
on the vortex.

The observation of ``momentogenesis'' in $^3$He confirms 
``baryogenesis'' in the electroweak model. The experiments, however, do 
not say anything about the cosmological process of baryogenesis since these
depend on various other cosmological factors such as departures from
thermal equilibrium and CP violation.

\section{Outlook}

In the study of the early universe, the last several decades 
have seen a remarkable confluence of ideas originating in 
vastly different branches of physics. Who could have imagined 
the possibility of fossils from the early universe and that 
one day we would be ``digging'' for them? That the mysteries
of the atom could be revealed by astronomical observations, while 
the secrets of the big bang may be locked in particle accelerators? 
It requires an even further stretch of imagination to contemplate
simulating the early universe in a vial of helium.
Yet this is the current state of early universe cosmology
and we can be sure of many equally surprising developments
in the years to come.

\smallskip

\noindent {\it Acknowledgements:} I am grateful to Andrew de Laix 
for the figures showing gravitational lensing by strings, to Martin 
White for providing the CMBR figures, and to the Department of Energy 
(USA) for research support.

\vskip 3 truein

Tanmay Vachaspati is Warren E. Rupp Associate Professor
of Science and Engineering at Case Western Reserve
University (Cleveland, USA). He obtained his Ph.D. 
under the supervision of Alexander Vilenkin 
at Tufts University in 1985. Since then he has been a postdoctoral
fellow at Bartol Research Institute (University of Delaware) and
D.A.M.T.P. (University of Cambridge) and on the faculty at
Tufts University. He was a Rosenbaum Fellow at the Isaac Newton
Institute in 1994, following which he joined Case Western Reserve
University as Associate Professor in 1995. 

Tanmay Vachaspati enjoys research in cosmology particularly
for its richness in diverse problems. His investigations of the 
early universe are stimulated by observations in astronomy and
cosmology, developments in theoretical particle physics, gravitational 
phenomena and condensed matter experiments. A hope that spurs
his research activities is the possibility that one day we will
directly probe the universe within the first second of its existence.


\begin{thebibliography}{999}

\bibitem{rebbisoliani} Some of the literature on this subject may
be found in the collection of reprints ``Solitons and Particles'',
C. Rebbi and G. Soliani (World Scientific, 1984).

\bibitem{tvdual} T. Vachaspati, Phys. Rev. Lett. {\bf 76}, 188 (1996).

\bibitem{abraham} M. Abraham, Phys. Zeitschrift {\bf 5}, 576 (1904).

\bibitem{lorentz} H. A. Lorentz, ``The Theory of Electrons'',
(G. R. Stechert and Co., New York, 1923).

\bibitem{mgleiser} M. Gleiser, Contemporary Physics, in press (1997).

\bibitem{avps} A. Vilenkin and E. P. S. Shellard, ``Cosmic
Strings and Other Topological Defects'', Cambridge University
Press (1994).

\bibitem{mhtk} M. B. Hindmarsh and T. W. B. Kibble, Rep. Prog. Phys.
{\bf 58}, 477 (1995).

\bibitem{avpr} A. Vilenkin, Phys. Rep. {\bf 121}, 263 (1985).

\bibitem{tk} T. W. B. Kibble, J. Phys. {\bf A9}, 1387 (1976).

\bibitem{abrikosov} A. A. Abrikosov, Sov. Phys. JETP {\bf 5}, 1174 (1957).

\bibitem{hnpo} H. B. Nielsen and P. Olesen, Nucl. Phys. {\bf B61}, 45 (1973).

\bibitem{th} G. `t Hooft, Nucl. Phys. {\bf B79}, 276 (1974).

\bibitem{ap} A. M. Polyakov, JETP Lett. {\bf 20}, 194 (1974).

\bibitem{aatv} T. Vachaspati and A. Ach\'ucarro, Phys. Rev. {\bf D44},
3067 (1991).

\bibitem{nambu} Y. Nambu, Nucl. Phys. {\bf B130}, 505 (1977).

\bibitem{lcrs} S. A Shectman, S. D. Landy, A. Oemler, D. L. Tucker,
H. Lin, R. P. Kirshner and P. L. Schechter, Ap. J. {\bf 470}, 172 (1996).

\bibitem{cfa} V. de Lapparent, M. J. Geller and J. P. Huchra, Ap. J. {\bf 302},
L1 (1986); J. P. Huchra, M. J. Geller and H. G. Corwin, Jr., Ap. J. Supp.
{\bf 99}, 391 (1995).

\bibitem{firas} J. C. Mather et. al., Ap. J. {\bf 420}, 440 (1994).

\bibitem{cobeandmore} C. L. Bennett et. al., Ap. J. {\bf 464}, L1 (1996).
A recent summary may be found in: G. Smoot and D. Scott, astro-ph/9711069
(1997).

\bibitem{yzmk} Ya. B. Zeldovich and M. Yu. Khlopov, Phys. Lett. {\bf B79},
239 (1978).

\bibitem{jp} J. Preskill, Phys. Rev. Lett. {\bf 43}, 1365 (1979).

\bibitem{ag} A. H. Guth, Phys. Rev. {\bf D23}, 347 (1981).

\bibitem{linde} A. Linde, ``Particle Physics and Inflationary Cosmology'',
(Harwood, Chur, Switzerland).

\bibitem{proceed} ``The Formation and Evolution of Cosmic Strings'',
eds. G. W. Gibbons, S. W. Hawking and T. Vachaspati, Cambridge University
Press (1990).

\bibitem{mh} M. B. Hindmarsh, Phys. Rev. Lett. {\bf 77}, 4495 (1996).

\bibitem{adlktv} A. A. de Laix, L. M. Krauss and T. Vachaspati, 
Phys. Rev. Lett. {\bf 79}, 1968 (1997).

\bibitem{nkas} N. Kaiser and A. Stebbins, Nature {\bf 310}, 391 (1984).

\bibitem{ballenetal} B. Allen, R. R. Caldwell, S. Dodelson, L. Knox,
E. P. S. Shellard and A. Stebbins, Phys. Rev. Lett. {\bf 79}, 2624 (1997). 

\bibitem{aalbrechtetal} A. Albrecht, R. A. Battye and J. Robinson,
Phys. Rev. Lett. {\bf 79}, 4736 (1997).

\bibitem{yz} Ya. B. Zeldovich, M. N. R. A. S. {\bf 192}, 663 (1980).

\bibitem{avlss} A. Vilenkin, Phys. Rev. Lett. {\bf 46}, 1169 (1981);
Erratum: Phys. Rev. Lett. {\bf 46}, 1496 (1981).

\bibitem{jsav} J. Silk and A. Vilenkin, Phys. Rev. Lett. {\bf 53}, 1700 (1984).

\bibitem{asetal} A. Stebbins, S. Veeraraghavan, R. Brandenberger,
J. Silk and N. Turok, Ap. J. {\bf 322}, 1 (1987).

\bibitem{tvlss} T. Vachaspati, Phys. Rev. Lett. {\bf 57}, 1655 (1988).

\bibitem{tvav} T. Vachaspati and A. Vilenkin, Phys. Rev. Lett. {\bf 67},
1057 (1991).

\bibitem{rees} M. Rees, M. N. R. A. S. {\bf 222}, 3 (1986).

\bibitem{hoganrees} C. J. Hogan and M. J. Rees, Nature {\bf 311}, 109 (1984).

\bibitem{jtaylor} D. R. Stinebring, M. F. Ryba, J. H. Taylor and R. W. Romani,
Phys. Rev. Lett. {\bf 65}, 285 (1990).

\bibitem{turok} U-L. Pen, U. Seljak and N. Turok, Phys. Rev. Lett. {\bf 79},
1611 (1997).

\bibitem{spergel} W. Hu, D. N. Spergel and M. White, Phys. Rev. {\bf D55},
3288 (1997). 

\bibitem{durrer} R. Durrer, M. Kunz, C. Lineweaver and M. Sakellariadou,
Phys. Rev. Lett. {\bf 79}, 5198 (1997).

\bibitem{pbgsds} P. Bhattacharjee, Phys. Rev. {\bf D40}, 3968 (1989);
G. Sigl, D. N. Schramm and P. Bhattacharjee, Astropart. Phys. {\bf 2},
401 (1994).

\bibitem{tkay} A. J. Gill and T. W. B. Kibble, Phys. Rev. {\bf D50},
3660 (1994). 

\bibitem{avcosmicrays} V. Berezinsky and A. Vilenkin, Phys. Rev. Lett.
{\bf 79}, 5202 (1997).

\bibitem{zurek} W. H. Zurek, Nature {\bf 317}, 505 (1985).

\bibitem{robertanne} ``Formation and Interaction of Topological Defects'', 
eds. A-C. Davis and R. Brandenberger, NATO ASI Series B, Physics {\bf 349}
(Plenum Press, New York, 1995).

\bibitem{chuang} I. Chuang, R. Durrer, N. Turok and B. Yurke, 
Science {\bf 251}, 1336 (1990). 

\bibitem{ams} M. J. Bowick, L. Chandar, E. A. Schiff and A. M. Srivastava,
Science {\bf 263}, 943 (1994).

\bibitem{he4} P. C. Hendry, N. S. Lawson, R. A. M. Lee, 
P. V. E. McClintock and C. D. H. Williams, Nature {\bf 368}, 315 (1994).

\bibitem{grenoble} C. Bauerle, Yu. M. Bunkov, S. N. Fisher, H. Godfrin
and G. R. Pickett, Nature {\bf 382}, 332 (1996).

\bibitem{helsinki} V. M. H. Ruutu, V. B. Eltsov, A. J. Gill, T. W. B. Kibble,
M. Krusius, Yu. G. Makhlin, B. Placais, G. E. Volovik and W. Xu, Nature
{\bf 382}, 334 (1996).

\bibitem{agill} A. J. Gill, Contemporary Physics, in press (1997).

\bibitem{gvtv} G. E. Volovik and T. Vachaspati, Int. J. Mod. Phys.
{\bf B10}, 471 (1996).

\bibitem{baryo3he} T. D. C. Bevan, A. J. Manninen, J. B. Cook,
J. R. Hook, H. E. Hall, T. Vachaspati and G. E. Volovik, Nature
{\bf 386}, 689 (1997). 

\bibitem{ew} E. Witten, Nucl. Phys. {\bf B249}, 557 (1985).

\bibitem{tvgf} T. Vachaspati and G. B. Field, Phys. Rev. Lett. {\bf 73},
373 (1994); Erratum: Phys. Rev. Lett. {\bf 74} (1995).

\bibitem{jgtv} J. Garriga and T. Vachaspati, Nucl. Phys. {\bf B438},
161 (1995).

\end{thebibliography}
\end{document}